\documentclass{elsart}
\usepackage{amssymb}

\renewcommand{\d}{{\mathrm d}}
\renewcommand{\bar}[1]{\overline{#1}}

\usepackage{indentfirst}
\usepackage{psfig,color}
\usepackage{epsfig}
\usepackage{epsf}
\usepackage{graphicx}

\journal{Physics Letters B}

\begin{document}

\begin{frontmatter}
\title{Contribution of asymmetric strange-antistrange sea to the
Paschos-Wolfenstein relation} %$\sin^{2}\theta_{w}$}

\author[pku]{Yong Ding},
\author[ccastpku]{Bo-Qiang Ma\corauthref{cor}}
\corauth[cor]{Corresponding author.} \ead{mabq@phy.pku.edu.cn}
\address[pku]{Department of Physics, Peking University, Beijing 100871, China}
\address[ccastpku]{CCAST (World Laboratory), P.O.~Box 8730, Beijing
100080, China\\
Department of Physics, Peking University, Beijing 100871, China}

%%To Editor: Please dont change the institute style in the above part, especially the ordering of the institutions.

\begin{abstract}
The NuTeV Collaboration reported a value of $\sin^{2}\theta_{w}$
measured in neutrino-nucleon deep inelastic scattering, and found
that the value is three standard deviations from the standard
model prediction. This result is obtained under the assumption
that the strange-antistrange sea quarks of nucleons are symmetric,
and that the up and down quark distributions are symmetric with
the simultaneous interchange of $u$$\leftrightarrow$$d$ and
$p$$\leftrightarrow$$n$. We discuss the contribution of asymmetric
strange-antistrange sea to the Paschos-Wolfenstein relation in the
extraction of weak mixing angle $\sin^{2}\theta_{w}$. We also
point out that the contribution of asymmetric strange-antistrange
sea should remove roughly 30--80\% of the discrepancy between the
NuTeV result and other determinations of $\sin^{2}\theta_{w}$,
when using the light-cone meson-baryon model to calculate the
contribution of the strange-antistrange sea.
\end{abstract}

\end{frontmatter}

\section{Introduction}

It is widely believed that the standard model is a low energy
remnant of some more fundamental theory. In the standard model,
the weak mixing angle $\sin^{2}\theta_{w}$ is one of the basic
quantities. The precise determination of $\sin^{2}\theta_{w}$
plays a key role in testing the standard model of electroweak
interaction. Its present value was consistent with all the known
electroweak observables \cite{abba}, until the NuTeV Collaboration
reported a value of $\sin^{2}\theta_{w}$ measured in
neutrino-nucleon deep inelastic scattering (DIS) with both
neutrino and antineutrino beams. The value \cite{zell}
$$\sin^{2}\theta_{w}=0.2277\pm0.0013~(\mbox{stat})\pm0.0009~(\mbox{syst}),$$
 which is three standard deviations larger than the
 value, $$\sin^{2}\theta_{w}=0.2227\pm0.0004,$$ measured in other
 electroweak processes \cite{abba}. Various source of systematic errors have
 been clearly identified and examined.
 For extracting $\sin^{2}\theta_{w}$, the NuTeV Collaboration
 measured the ratio of neutrino neutral-current and charge-current
 cross sections on iron \cite{zell}. This procedure is closely related to
 the Paschos-Wolfenstein relation \cite{pash}:
\begin{equation}
R^{-}=\frac{\sigma^{{\nu}N}_{NC}-\sigma^{\overline{\nu}N}_{NC}}{\sigma^{{\nu}N}_{CC}
-\sigma^{\overline{\nu}N}_{CC}}=\frac{1}{2}-\sin^{2}\theta_{w}.\label{ratio}
\end{equation}
Because the NuTeV Collaboration did not strictly measure the
Paschos-Wolfenstein relation, Eq.~(\ref{ratio}), there are a
number of corrections that need to be considered, such as charge
symmetry violation \cite{LT}, which should reduce roughly
one-third of the discrepancy between the NuTeV result and all
accepted average value of $\sin^{2}\theta_{w}$, nuclear effect,
which arises from the higher twist effect of nuclear shadowing
\cite{blw}, neutron excess \cite{kuma}, although such modification
are not measured, differences in shadowing from photons, $W^{\pm}$
and $Z^{0}$s \cite{MT}, asymmetry in the $s$ and $\bar{s}$
distributions \cite{CS}, nuclear correction, discussed in
Ref.~\cite{KSY} by noting nuclear modification of $F_{2}$, also
recently QCD correction \cite{kret}, and so on. In addition, the
discussion of possible uncertainties and physics behind the
anomaly can be found in Ref.~\cite{dav}.

Eq.~(\ref{ratio}) is based on the assumption of symmetric quark
and antiquark distributions in the nucleon sea. In fact, the study
of the quark sea in the nucleon is important to understand the
nucleon structure and the strong interaction. Usually, we assume
that the quark and antiquark sea are symmetric, but we should note
that it may have asymmetry to some extent \cite{BM}. It is rather
difficult to study the asymmetry of the up and down sea in
experiment, because we hardly can distinguish the up and down sea
quarks from the corresponding valence quarks in the nucleon bound
state. However, for the strange quark sea, it is relatively
accessible and there have been analyses of experimental data
\cite{sbr,arne,blt,bpz}, which suggest the asymmetry of $s$ and
$\bar{s}$ distributions in the nucleon sea. Also, there are some
theoretical discussions on this issue \cite{BM,ST,bw,hss,cm,caos}.
Brodsky and Ma \cite{BM} proposed a light-cone meson-baryon
fluctuation model to describe the $s$-$\bar{s}$ distributions and
found that $s<\bar{s}$ in small $x$ region and $s>\bar{s}$ in
large $x$ region. A significantly different conclusion was
obtained by Holtmann, Szczurek and Speth \cite{hss} from
Ref.~\cite{BM} by using the meson cloud model with fluctuation
function \cite{ST,hss}. Cao and Signal \cite{caos} obtained a
phenomenological analysis of $s$-$\bar{s}$ asymmetry in the
nucleon sea when using two different models: light-cone model
\cite{BM} and meson cloud model \cite{ST,caos}. In this paper, we
consider the role of the $s$-$\bar{s}$ asymmetry in the nucleon
sea by using the light-cone meson-baryon fluctuation model
\cite{BM}, and find that it should account for roughly 30--80\% of
the discrepancy between the NuTeV result and other accepted value
of $\sin^{2}\theta_{w}$. Our result is different from the previous
conclusion \cite{CS} that the effect of asymmetric
strange-antistrange sea is fairly small and does not affect the
NuTeV extraction of $\sin^{2}\theta_{w}$.

\section{Modified Paschos-Wolfenstein Relation}

The Paschos-Wolfenstein relation was derived for $s(x)=\bar{s}(x)$
in the nucleon sea. In this section, we shall derive a revised
expression for $s(x)\neq\bar{s}(x)$. The cross sections for
neutrino- and antineutrino-nucleon neutral current interaction
have the form \cite{londt}
\begin{eqnarray}
\frac{\d^{2}\sigma^{\nu(\bar{\nu})}_{NC}}{\d x\d y}&=&\pi
s\left(\frac{\alpha}{2\sin^{2}\theta_{w}\cos^{2}\theta_{w}M^{2}_{Z}}
\right)^{2}(\frac{M^{2}_{Z}}{M^{2}_{Z}+Q^{2}})^{2} [xyF^{Z}_{1}(x,Q^{2}) \nonumber\\
&&
+(1-y-\frac{xym^{2}_{N}}{s})F^{Z}_{2}(x,Q^{2})\pm(y-\frac{y^{2}}{2})xF^{Z}_{3}(x,Q^{2})],
\end{eqnarray}
and the cross sections for neutrino- and antineutrino-nucleon
charge current reaction have the form \cite{londt}
\begin{eqnarray}
\frac{\d^{2}\sigma^{\nu(\bar{\nu})}_{CC}}{\d x \d
y}&=&{\pi}s\left(\frac{\alpha}{2\sin^{2}\theta_{w}
M^{2}_{W}}\right)^{2}(\frac{M^{2}_{W}}{M^{2}_{W}+Q^{2}})^{2}
 [xyF^{W^{\pm}}_{1}(x,Q^{2}) \nonumber\\
&& +(1-y-\frac{xym^{2}_{N}}{s})
F^{W^{\pm}}_{2}(x,Q^{2})\pm(y-\frac{y^{2}}{2})xF^{W^{\pm}}_{3}(x,Q^{2})
],
\end{eqnarray}
where $Q^{2}=-q^{2}$ is the square of the four momentum transfer
for the reaction, $M_{W}(M_{Z})$ is the mass of the charge
(neutral) current interacting weak vector boson, $\theta_{w}$ is
the Weinberg angle, and $x=Q^{2}/2p\cdot q$, $y=p\cdot q/p\cdot
k$, and $s=(k+p)^{2}$ are the DIS variables for four momentum $k$
($p$) of the initial state neutrino or antineutrino (nucleon). The
structure functions $F^{W^{\pm} p}_{i}(x,Q^{2})$ on proton ($p$),
which only depend on $x$, as $Q^{2}\rightarrow\infty$, were given
by \cite{londt}
\begin{eqnarray}
\lim_{Q^{2}\rightarrow\infty}F^{W^{+}p}_{1}(x,Q^{2})&=&d^{p}(x)+\bar{u}^{p}(x)+s^{p}(x)+\bar{c}^{p}(x),\nonumber\\
\lim_{Q^{2}\rightarrow\infty}F^{W^{-}p}_{1}(x,Q^{2})&=&u^{p}(x)+\bar{d}^{p}(x)+\bar{s}^{p}(x)+c^{p}(x),\nonumber\\
\frac{1}{2}\lim_{Q^{2}\rightarrow\infty}F^{W^{+}p}_{3}(x,Q^{2})&=&d^{p}(x)-\bar{u}^{p}(x)+s^{p}(x)-\bar{c}^{p}(x),\nonumber\\
\frac{1}{2}\lim_{Q^{2}\rightarrow\infty}F^{W^{-}p}_{3}(x,Q^{2})&=&u^{p}(x)-\bar{d}^{p}(x)-\bar{s}^{p}(x)+c^{p}(x),\nonumber\\
F^{W^{\pm}p}_{2}(x,Q^{2})&=&2xF^{W^{\pm}p}_{1}(x,Q^{2}).\label{stru}
\end{eqnarray}
The structure functions of neutral current reaction take the form
\cite{londt}
\begin{eqnarray}
\lim_{Q^{2}\rightarrow\infty}F^{Zp}_{1}(x,Q^{2})&=&\frac{1}{2}\left[(u^{2}_{V}+u^{2}_{A})\left(u^{p}(x)+\bar{u}^{p}(x)
    +c^{p}(x)+\bar{c}^{p}(x)\right)\right.\nonumber\\&&\left.+(d^{2}_V+d^{2}_{A})\left(d^{p}(x)+\bar{d}^{p}(x)
    +s^{p}(x)+\bar{s}^{p}(x)\right) \right],\nonumber\\
\lim_{Q^{2}\rightarrow\infty}F^{Zp}_{3}(x,Q^{2})&=&2\left[u_{V}u_{A}\left(u^{p}(x)-
    \bar{u}^{p}(x)+c^{p}(x)-\bar{c}^{p}(x)\right)\right.\nonumber\\&&\left.+d_{V}d_{A}\left(d^{p}(x)-
    \bar{d}^{p}(x)+s^{p}(x)-\bar{s}^{p}(x)\right)\right], \nonumber\\
F^{Zp}_{2}(x,Q^{2})&=&2xF^{Zp}_{1}(x,Q^{2}),\label{struc}
\end{eqnarray}
and the corresponding structure functions
$F^{W^{\pm}(Z)}_{i}(x,Q^{2})$ for neutrons are given by replacing
superscripts $p\rightarrow$$n$ in Eqs.~(\ref{stru}),
(\ref{struc}), with the assumption of charge symmetry for parton
distributions
\begin{eqnarray}
   d^{n}(x)&=&u^{p}(x),\nonumber\\
   u^{n}(x)&=&d^{p}(x),\nonumber\\
  s^{n}(x)&=&s^{p}(x)=s(x),\nonumber\\
  c^{n}(x)&=&c^{p}(x)=c(x).\label{pdis}
\end{eqnarray}
In Eqs.~(\ref{struc}), $u_{V}$, $d_{V}$, $u_{A}$ and $d_{A}$ are
vector and axial-vector couplings:
$$u_{V}=\frac{1}{2}-\frac{4}{3}\sin^{2}\theta_{w},  \ \ u_{A}=\frac{1}{2},$$
$$d_{V}=-\frac{1}{2}+\frac{2}{3}\sin^{2}\theta_{w}, \ \ d_{A}=-\frac{1}{2}.$$
Using these equations, we obtain the modified Paschos-Wolfenstein
relation:
\begin{eqnarray}
  R^{-}_{N}=\frac{\sigma^{\nu N}_{NC}-\sigma^{\bar{\nu}N}_{NC}}{\sigma^{\nu N}_{CC}-\sigma^{\bar{\nu}N}_{CC}}
= R^{-}-\delta R^{-}_{s}.\label{correction}
\end{eqnarray}
Here, $\delta R^{-}_{s}$ is the correction to the
Paschos-Wolfenstein relation $R^{-}$ from the asymmetry of
$s$-$\bar{s}$ distribution in the nucleon sea,
\begin{eqnarray}
  \delta
  R^{-}_{s}=-(-1+\frac{7}{3}\sin^{2}\theta_{w})\frac{S^{-}}{Q_V+3 S^{-}},\label{rs}
\end{eqnarray}
where $Q_V \equiv\int^{1}_{0} x[u_{V}(x)+d_{V}(x)]\d x$ and
$S^{-}\equiv\int^{1}_{0} x[s(x)-\bar{s}(x)]\d x$. During this
procedure of getting $R^{-}_{N}$, we assume isospin symmetry and
$c(x)=\bar{c}(x)$. In this way, we obtain the correction $\delta
R^{-}_{s}$ and below we shall calculate $S^{-}$ and $Q_V$ by using
the light-cone two-body wave function model \cite{BM} and the
light-cone spectator model \cite{ma}.

\section{Strange-Antistrange Asymmetry }

We shall adapt the light-cone two-body wave function model
\cite{BM} to calculate $S^{-}$. In this light-cone formalism
\cite{ck}, the hadronic wave function can be expressed by a series
of light-cone wave functions multiplied by the Fock states, for
example, the proton wave function can be written as
\begin{eqnarray}
   \left|p\right\rangle=\left|uud\right\rangle\Psi_{uud/p}+ \left|uudg\right\rangle\Psi_{uudg/p}+\sum_{q\bar{q}}
   \left|uudq \bar{q}\right\rangle\Psi_{uudq\bar{q}/p}+\cdots.
\end{eqnarray}
Brodsky and Ma made an approximation \cite{BM}, which suggests
that the intrinsic sea part of the proton function can be
expressed as a sum of meson-baryon Fock states. For example:
$P(uuds\bar{s})=K^{+}(u\bar{s})+\Lambda(uds)$ for the intrinsic
strange sea, the higher Fock states are less important, the $ud$
in $\Lambda$ serves as a spectator in the quark-spectator model
\cite{ma}, for which we choose
\begin{eqnarray}
   \Psi_{D}(x,\mathbf{k}_{\perp})=A_{D}\exp(-M^{2}/8\alpha^{2}_{D}),\label{si}
\end{eqnarray}
\begin{eqnarray}
 \Psi_{D}(x,\mathbf{k}_{\perp})=A_{D}(1+M^{2}/\alpha^{2}_{D})^{-P},\label{psi}
 \end{eqnarray}
where $\Psi_{D}(x,\mathbf{k}_{\perp})$, is a two-body wave
function which is a function of invariant masses for meson-baryon
state:
\begin{eqnarray}
  M^{2}=\frac{m^{2}_{1}+\mathbf{k}^{2}_{\bot}}{x}+\frac{m^{2}_{2}+\mathbf{k}^{2}_{\bot}}{1-x},
\end{eqnarray}
where $\mathbf{k}_{\perp}$ is the initial quark transversal
momentum, $m_{1}$ and $m_{2}$ are the masses for quark $q$ and
spectator $D$, $\alpha_{D}$ sets the characteristic internal
momentum scale, and $P$ is the power constant which is chosen as
$P=3.5$ here. The momentum distribution of the intrinsic $s$ and
$\bar{s}$ in the $K^{+}\Lambda$ state can be modelled from the
two-level convolution formula:
\begin{eqnarray}
   s(x)&=&\int^{1}_{x}\frac{\d y}{y}f_{\Lambda/K^{+}\Lambda}(y)q_{s/\Lambda}(x/y),\nonumber\\
   \bar{s}(x)&=&\int^{1}_{x}\frac{\d y}{y}f_{K^{+}/K^{+}\Lambda}(y)q_{\bar{s}/K^{+}}(x/y),
\end{eqnarray}
where $f_{\Lambda/K^{+}\Lambda}(y)$, $f_{K^{+}/K^{+}\Lambda}(y)$
are the probabilities of finding $\Lambda, K^{+}$ in the
$K^{+}\Lambda$ state with the light-cone momentum fraction $y$,
for the Gaussian type:
\begin{eqnarray}
  f_{\Lambda/K^{+}\Lambda}(y)&=&\int^{+\infty}_{-\infty}\d\mathbf{k}_{\bot}\bigg{|}A_{D}
  \exp[-\frac{1}{8\alpha^{2}_{D}}(\frac{m^{2}_{\Lambda}+\mathbf{k}^{2}_{\bot}}{y}+\frac{m^{2}_{K^{+}}+\mathbf{k}^{2}_{\bot}}{1-y})]\bigg{|}^{2},\nonumber\\
 f_{K^{+}/K^{+}\Lambda}(y)&=&\int^{+\infty}_{-\infty}\d\mathbf{k}_{\bot}\bigg{|}A_{D}
  \exp[-\frac{1}{8\alpha^{2}_{D}}(\frac{m^{2}_{K^{+}}+\mathbf{k}^{2}_{\bot}}{y}+\frac{m^{2}_{\Lambda}+\mathbf{k}^{2}_{\bot}}{1-y})]\bigg{|}^{2},
\end{eqnarray}
and $q_{s/\Lambda}(x/y)$, $q_{\bar{s}/K^{+}}(x/y)$ are the
probabilities of finding $s$, $\bar{s}$ quarks in $\Lambda, K^{+}$
state with the light-cone momentum fraction $x/y$, for the
Gaussian type:
\begin{eqnarray}
  q_{s/\Lambda}(x/y)&=&\int^{+\infty}_{-\infty}\d\mathbf{k}_{\bot}\bigg{|}A_{D}
  \exp[-\frac{1}{8\alpha^{2}_{D}}(\frac{m^{2}_{s}+\mathbf{k}^{2}_{\bot}}{x/y}+\frac{m^{2}_{D}+\mathbf{k}^{2}_{\bot}}{1-x/y})]\bigg{|}^{2},\nonumber\\
  q_{\bar{s}/K^{+}}(x/y)&=&\int^{+\infty}_{-\infty}\d\mathbf{k}_{\bot}\bigg{|}A_{D}
  \exp[-\frac{1}{8\alpha^{2}_{D}}(\frac{m^{2}_{\bar{s}}+\mathbf{k}^{2}_{\bot}}{x/y}+\frac{m^{2}_{q}+\mathbf{k}^{2}_{\bot}}{1-x/y})]\bigg{|}^{2}.
\end{eqnarray}
Two wave function models, the Gaussian type and the power-law
type, are adopted \cite{BM} to evaluate the asymmetry of
strange-antistrange sea, and almost identical distributions of
$s$-$\bar{s}$ are obtained in the nucleon sea. In this paper, we
also consider the two types of wave functions, Eqs.~(\ref{si}) and
(\ref{psi}).

\begin{figure}%[htb]
\begin{center}
\includegraphics[width=9.5cm]{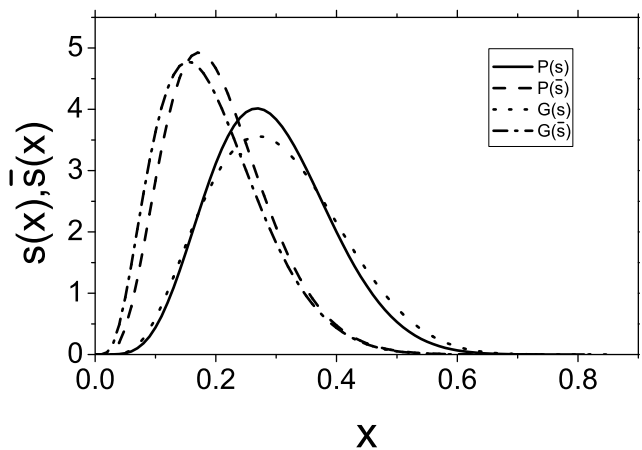}
\caption{\small Distributions for $s(x)$ and $\bar{s}(x)$. $P(s)$
($G(s)$) is the $s$ distribution with the power-law wave function
(the Gaussian wave function) and $P(\bar{s}$) ($G(\bar{s}$)) is
the $\bar s$ distribution with the power-law wave function (the
Gaussian wave function).}\label{ssbar}
\end{center}
\end{figure}

\begin{figure}%[htb]
\begin{center}
\includegraphics[width=9.5cm]{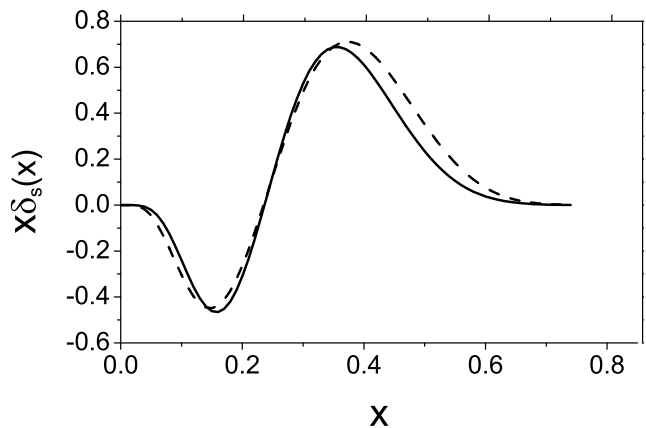}
\caption{\small Distributions for $x\delta_{s}(x)$, with
$\delta_{s}(x)$=$s(x)-\bar{s}(x)$. The solid curve is for the
power-law wave function and the dashed curve is for the Gaussian
wave function.}\label{dssbar}
\end{center}
\end{figure}

The up and down valence quark distributions in the proton are
calculated by using the quark-diquark model. The unpolarized
valence quark distribution in the proton is \cite{ma}
\begin{eqnarray}
   u_{V}(x)&=&\frac{1}{2}a_{S}(x)+\frac{1}{6}a_{V}(x),\nonumber\\
   d_{V}(x)&=&\frac{1}{3}a_{V}(x),
\end{eqnarray}
where $a_{D}(x)$ ($D=S$ or $V$, with $S$ standing for scalar
diquark Fock state and $V$ standing for vector diquark state)
denotes that the amplitude for the quark $q$ is scattered while
the spectator is in diquark state $D$ \cite{kbb}, and can be
written as:
\begin{eqnarray}
   a_{D}(x)\propto\int[\d\mathbf{k}_{\bot}]\bigg{|}\Psi_{D}(x,\mathbf{k}_{\bot})\bigg{|}^{2}.
\end{eqnarray}
The values of parameters $\alpha_{D}$, $m_{q}$, and $ m_{D}$ can
be adjusted by fitting the hardonic properties. For light-flavor
quarks, we simple choose $m_{q}=330$~MeV, $\alpha_{D}=330$~MeV,
$m_{S}=600$~MeV, $m_{V}=900$~MeV and $m_{s}=m_{\bar{s}}=480$~MeV
\cite{BM}. Because the fluctuation functions were normalized to 1
in Ref.~\cite{BM}, we can obtain the different distributions for
$s$ and $\bar{s}$ in the nucleon. In the same way, we can get the
distributions of the up and down valence quarks, for which the
integrated amplitude $\int_0^1 \d x \,a_D(x)$ must be normalized
to 3 in a spectator model \cite{ma,ma2}. Assuming isospin
symmetry, we can get the valence distributions in the nucleon
which implies $N=(p+n)/2$
\begin{eqnarray}
  u^{N}_{V}(x)=\frac{1}{2}\left[\frac{1}{2}a_{S}(x)+\frac{1}{2}a_{V}(x)\right],\nonumber\\
  d^{N}_{V}(x)=\frac{1}{2}\left[\frac{1}{2}a_{S}(x)+\frac{1}{2}a_{V}(x)\right].
\end{eqnarray}
Thus, using this model, we obtain the distributions of $s$ and
$\bar{s}$ in the nucleon sea. The numerical result is given in
Fig.~\ref{ssbar}. One can find that $s<\bar{s}$ as $x<0.235$,
$s>\bar{s}$ as $x>0.235$, this result is opposite to the
prediction from the meson cloud model \cite{CS}. Similarly, one
can obtain the shape of $x(s-\bar{s})$ in Fig.~\ref{dssbar}. From
Eq.~(\ref{correction}), one can find that a shift of $\delta
R^{-}_{s}$ should lead to a shift in the $R^{-}$, which affect the
extraction of $\sin^{2}\theta_{w}$, Eq.~(\ref{rs}). The result of
our calculation is 0.0042$<S^{-}<$0.0106 (0.0035$<S^{-}<$0.0087)
for the Gaussian wave function (for the power-law wave function),
which corresponds to $P_{K^{+}\Lambda}$=4\%, 10\%. Hence,
$0.0017<\delta R^{-}_{s}<0.0041$ ($0.0014<\delta
R^{-}_{s}<0.0034$), for the Gaussian wave function (the power-law
wave function). The shift in $\sin^{2}\theta_{w}$ can reduce the
discrepancy from 0.005 to 0.0033 (0.0036) ($P_{K^{+}\Lambda}$=4\%)
or 0.0009 (0.0016) ($P_{K^{+}\Lambda}$=10\%).

\section{Summary}

Intrinsic sea quarks play a crucial role in understanding the
structure of the nucleon and strong interaction, such as the
effect of the strange and antistrange quark distributions to the
nucleon structure. In this work, we have re-examined the asymmetry
of $s$-$\bar{s}$ distribution in the nucleon with the light-cone
meson-baryon model. Considering this asymmetry, we derived a
modified Paschos-Wolfeinstein relation. Though there have been
evidences for the asymmetry of $s$-$\bar{s}$ distribution in the
nucleon sea suggested by analyses \cite{sbr,arne,blt,bpz}, this
asymmetry need to be directly confirmed experimentally. We have
strong theoretical arguments about the sign and magnitude of the
correction to the Paschos-Wolfeinstein relation. In particular,
this correction should make a significant contribution to the
NuTeV extraction of the weak mixing angle $\sin^{2}\theta_{w}$ by
a deviation 30--80\%, which is corresponding to the assumption
that the probability is 4--10\% for the $K^{+}\Lambda$ state.
Therefore it is important to investigate the effect of asymmetric
strange-antistrange sea more carefully in future experiments.

{\bf Acknowledgments }

This work is partially supported by National Natural Science
Foundation of China under Grant Numbers 10025523 and 90103007.


\begin{thebibliography}{99}
\bibitem{abba} See, e.g., D. Abbaneo, $et. al.$, hep-ex/0112021.
\bibitem{zell} G.P. Zeller, $et. al.$, Phys. Rev. Lett. {\bf 88}
(2002) 091802.
\bibitem{pash} E.A. Paschos, L. Wolfenstein, Phys. Rev. D {\bf 7} (1973)
91.
\bibitem{LT} J.T. Londergan, A.W. Thomas, Phys. Lett. B {\bf
558} (2003) 132.
\bibitem{blw} C. Boros, J.T. Londergan, A.W. Thomas, Phys.
Rev. D {\bf 59} (1999) 074021;

Phys. Rev. D {\bf 58} (1998) 114030.
\bibitem{kuma} S. Kumano, Phys. Rev. D {\bf 66} (2002) 111301.
\bibitem{MT} G.A. Miller, A.W. Thomas, hep-ex/0204007.
\bibitem{CS} F.-G. Cao, A.I. Signal, Phys. Lett. B {\bf 559} (2003)
229.
\bibitem{KSY} S. Kovalenko, I. Schmidt, J.-J. Yang, Phys. Lett.
B {\bf 546} (2002) 68.
\bibitem{kret} S. Kretzer, $et. al.$, hep-ph/0312322.
\bibitem{dav} S. Davidson, $et. al.$, JHEP {\bf 02} (2002) 037.

\bibitem{BM} S.J. Brodsky, B.-Q. Ma, Phys, Lett. B {\bf 381}
(1996) 317.

\bibitem{sbr} W.G. Seligman, $et. al.$, Phys. Rev. Lett. {\bf 79} (1997)
1213;

A.O. Bazarko, $et. al.$, Z. Phys, C {\bf 65} (1995) 189;

S.A. Rabinowitz, $et. al.$, Phys. Rev. Lett. {\bf 70} (1993) 134.
\bibitem{arne} M. Arneodo, $et. al.$, Nucl. Phys. B {\bf 483} (1997) 3.
\bibitem{blt} C. Boros, J.T. Londergan, A.W. Thomas, Phys.
Rev. Lett. {\bf 81} (1998) 4075.
\bibitem{bpz} V. Barone, C. Pascaud, F. Zomer, Eur. Phys. J. C {\bf
12} (2000) 243.

\bibitem{ST} A.I. Signal, A.W. Thomas, Phys. Lett. B {\bf 191} (1987)
205 .
\bibitem{bw} M. Burkardt, J. Warr, Phys. Rev. D {\bf 45} (1992)
958.
\bibitem{hss} H. Holtmann, A. Szczurek, J. Speth, Phys.
A {\bf 569} (1996) 631.
\bibitem{cm} H.R. Christiansen, J. Magnin, Phys. Lett. B {\bf 445} (1998)
8.
\bibitem{caos} F.-G. Cao, A.I. Signal, Phys. Rev. D {\bf 60} (1999)
074021.
\bibitem{londt} J.T. Londergan, A.W. Thomas, Prog. Part. Nucl. Phys. {\bf 41} (1998) 49.
\bibitem{ma} B.-Q. Ma, Phys, Lett. B {\bf 375} (1996)
320.
\bibitem{ck} R. Carlitz, Phys. Lett. B {\bf 58} (1975) 345 ;

J. Kaur, Nucl. Phys. B {\bf 128} (1977) 219.
\bibitem{kbb} J.B. Kogut, D.E. Soper, Phys. Rev. D {\bf 1} (1970)
2901;

J.B. Bjorken, J.B. Kogut, D.E. Soper, $ibid.$ {\bf 3} (1971) 1328;

S.J. Brodsky, R. Roskies, R. Suaya, $ibid.$ {\bf 8} (1973) 4574.
\bibitem{ma2}
B.-Q. Ma, D. Qing, and I. Schmidt, Phys. Rev. C {\bf 65} (2002)
035205.

\end{thebibliography}
\end{document}